\documentclass[aip,jap,reprint,superscriptaddress]{revtex4-2}
\usepackage{amsmath}
\usepackage{xcolor}
\usepackage{graphicx}
\usepackage{lineno}
\input{epsf}
\begin{document}
\title{Stability of thermally bistable states and their switching in superconducting weak link}
\author{Sourav Biswas}
\email[]{sbiswas.physics@gmail.com}
\affiliation{Department of Physics, Indian Institute of Technology Kanpur, Kanpur 208016, India}
\affiliation{Braun Center for Submicron Research, Department of Condensed Matter Physics, Weizmann Institute of Science, Rehovot 7610001, Israel}
\author{Pankaj Wahi}
\affiliation{Department of Mechanical Engineering, Indian Institute of Technology Kanpur, Kanpur 208016, India}
\author{Anjan Kumar Gupta}
\affiliation{Department of Physics, Indian Institute of Technology Kanpur, Kanpur 208016, India}

\begin{abstract}
Superconducting weak link (WL), acting as a Josephson junction (JJ), is one of the widely used elements in superconductor science and quantum circuits. A hysteretic JJ with robust switching between its superconducting and resistive state is an excellent candidate for single-photon detection. However, the ubiquitous fluctuations in the junction strongly influence the stability of the states and, thus, the transition from one to the other. Here, we present an experimental study of switching statistics of critical and retrapping currents of a JJ based on niobium WL in its hysteretic regime. The mean lifetimes of the two metastable states, namely, the zero-voltage superconducting state and finite-voltage resistive state, are estimated from the distributions. Further, close to the hysteresis crossover temperature, observed telegraphic noise in the time domain due to random switching between the states provides their lifetimes directly. We present a thermal model introducing a double-well (bistable) feature with an intriguing quantity with respect to the devices' temperature states. The effects of temperature fluctuations on the stability of the states are shown. We discuss our results toward further improvement of the efficiency of superconducting WL or nanowire single-photon detectors.

\end{abstract}

\maketitle
\section{Introduction}

Josephson junctions (JJs) have been of extensive experimental and theoretical research interest for their implications in many systems. For instance, a hysteretic JJ 
is employed in microwave single-photon detection \cite{detector1,detector2,detector3,detector4,singlephoton-1,singlephoton-2,pankra-2,nanowire-detector,first-snspd,spd,spd1,spd2}, a non-linear JJ acts as a quantum bit system \cite{qbit1,qbit2}, and two JJs forming a superconducting quantum interference device (SQUID) can probe magnetism at nanoscale \cite{mag1,mag2}. In a current biased hysteretic JJ, the junction switches from the zero-voltage superconducting state to finite-voltage resistive state at a critical current $I_{\rm c}$ during the current ramp-up, while it comes back to the superconducting state at a retrapping current $I_{\rm r}$ ($< I_{\rm c}$). According to the resistively and capacitively shunted junction (RCSJ) model, junctions' characteristics are described by the dynamics of the superconducting phase ($\varphi$) in a tilted washboard potential \cite{tinkham book,squidbook}. The junction capacitance measuring the drag in the potential decides the retrapping and, thus, the hysteresis. The sharp jump at the critical current of a hysteretic JJ makes it a prominent tool to detect single-photon.

A practical JJ suffers inevitable thermal or quantum fluctuations that lead the phase to change across the junction. This is known as the phase-slip process, \cite{tinkham book,phaseslip, phaseslip-1} which causes the junction to transit (retrap) at a current different than the intrinsic $I_{\rm c}$ ($I_{\rm r}$) value, giving rise to a spread. The first detailed study of the phase-slip induced switching and distribution in $I_{\rm c}$ of a superconductor--insulator--superconductor (SIS) JJ was performed by Fulton and Dunkleberger \cite{fulton}. Later, several works were carried out to investigate both $I_{\rm c}$ and $I_{\rm r}$ statistics, their temperature dependence, decay of the metastable states in SIS and SNS (superconductor--normal metal--superconductor) junctions \cite{prl2,prl1,tempswitch,prb1,prbrapid,jap,grapheneJJ-1,SNS-JJ,pankra-1}. While the understanding is mostly made within the traditional RCSJ model, the latter is not often competent for explaining $I_{\rm r}$ in JJs based on the superconducting weak link (WL), nanowire \cite{likharev,bezryadin-1,JJ-nanowire,moge-nanowire,pengli-prl}, and also in the SNS junctions \cite{herve}. Here the hysteresis is attributed to dominantly thermal due to Joule heating \cite{herve,scocpol,tinkham nano wire,hazra,anjanjap,sourav1}. For such a JJ with multiple adjacent leads, thermal instability of the normal-superconductor (NS) interface, arising from phase-slip and associated heating, determines $I_{\rm r}$ \cite{nayana,bezryadin-2,nikhil prl,nikhilsust}. With recently growing research attention on WL or nanowire, it is, thus, important to explore the effects of fluctuations, instability of the states to understand them as a two-state system, the states' transition, and to improve their performance as a detector. The role of fluctuation and dissipation is also crucial in understanding the coherence and optimizing such JJ devices for quantum applications \cite{hadfield}.

In this article, we study switching current distributions of $I_{\rm c}$ and $I_{\rm r}$ in the hysteretic regime of a niobium (Nb) WL having a large critical current. The distributions are used to estimate the lifetimes of the two metastable states, viz., superconducting state and resistive state. When $I_{\rm c}$ and $I_{\rm r}$ are nearly separated and their spreads overlap, we observe the direct bistable characteristics via a random telegraphic signal in voltage with time. A simple thermal model for a superconducting lead dictates the temperature profile, NS interface, and reveals the existence of a bistable thermal state above $I_{\rm r}$. We qualitatively discuss the thermal fluctuation effect on the stability of NS interface and, hence, its lifetime. A quantitative analysis of the timescale related to the stability of such a dissipative system needs further investigation.
\begin{figure}[h!]
\includegraphics[width=\columnwidth]{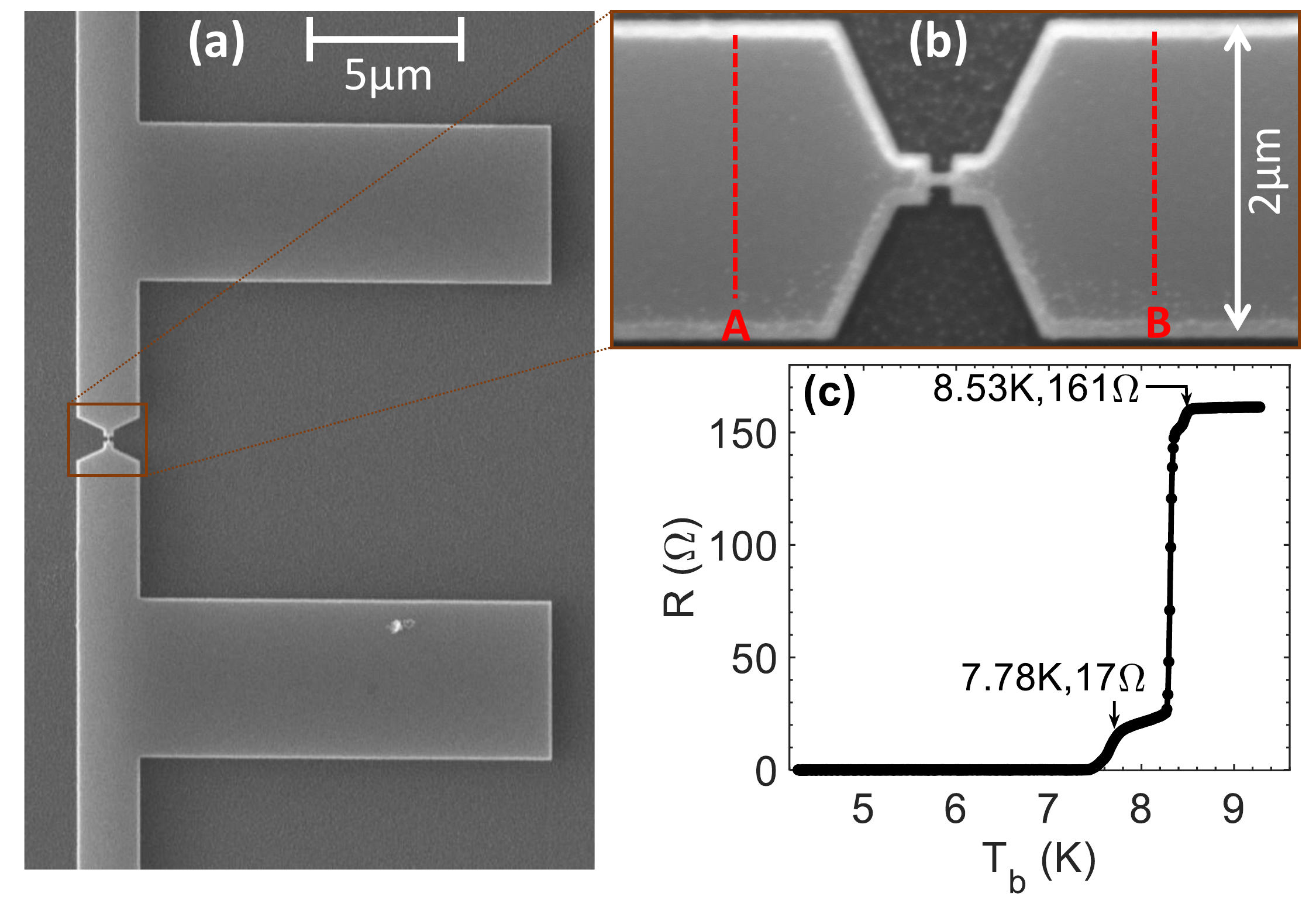}
	\caption{(a) SEM image of the weak link (WL) with various leads. (b) Magnified image of the WL along with the nearest leads. The length and the width of the link are 160 and 40nm, respectively. AB denotes the section of the device that becomes normal above $I_{\rm r}$ (see texts). (c) Resistance $R$ vs bath temperature $T_{\rm b}$, showing the critical temperature of $T_{\rm c}$ = 8.53K and a gradual diminishing of $R$ afterward. The second transition at 7.78K is due to the WL and nearest short leads.}
	\label{fig:1}
\end{figure} 

\section{Experimental details}
The devices were fabricated using 40nm thick Nb deposited on a Si substrate in an ultra-high vacuum chamber. Electron-beam and laser lithography were used to pattern the smaller (WL and narrow leads) structures and the bigger parts (wider leads and contact pads). An Al layer of 25nm was then deposited and lifted off. With Al as the protective mask, Nb was etched by SF$_6$ reactive-ion. Finally, removing Al by chemical etching, the Nb device structure was uncovered. Afterward, we cut Nb thickness down to $\sim$20($\pm2$)nm to reduce critical current. \mbox{Figures \ref{fig:1}(a) and \ref{fig:1}(b)} show the scanning electron micrograph (SEM) images of our WL geometry.

Electrical transport measurements were carried out with a homemade cryostat in liquid helium down to 4.2K. We deployed a specially designed sample holder containing copper powder to filter the high-frequency noise. The signal-carrying wires, connected from the device to the external electronics were adequately shielded, and each of them was passed through a commercial $\pi$-filter. A ground-isolated current source minimized the external noise. The data were recorded using a data acquisition card (capable of more than 200 kilo-sampling/s) and a LabView program. We performed the experiments on multiple devices that showed similar observations. The detailed results from one device are described here. 

\section{Measurement results and analysis}
The measurement starts with four-probe resistance ($R$) as a function of the bath temperature $T_{\rm b}$. Figure \ref{fig:1}(c) shows the onset superconducting transition of Nb at critical temperature $T_{\rm c}$ = 8.53K. At $T_{\rm c}$, $R$ drops from 161$\Omega$ to 24.7$\Omega$ and goes to zero following multiple steps. The estimated residual resistivity ratio, $R_{\rm 300K}/R_{\rm T_c}\approx3$ is a measure of the quality of our Nb film. The resistance of 161$\Omega$ at onset $T_{\rm c}$ leads to an estimate of the sheet resistance $R_{\rm sh}$ to be 4.48$\Omega$ (see Appendix-A). Thus, the WLs' normal resistance is $R_{\rm sh}\times L/w =$ 17.9$\Omega$, which is consistent with the observed second transition at 7.78K in the $R-T_{\rm b}$ plot [see Fig. \ref{fig:1}(c)].

\begin{figure}[h!]
	\includegraphics[width=\columnwidth]{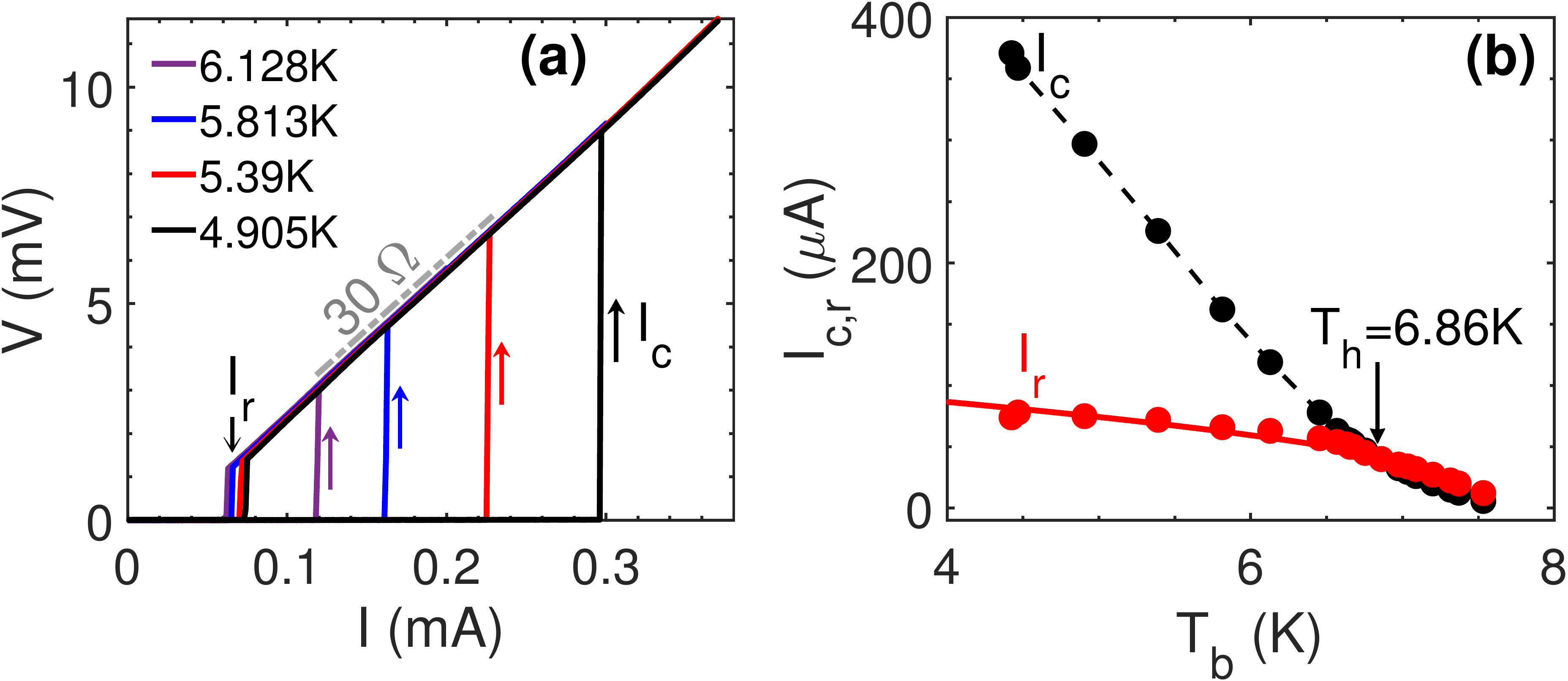}
	\caption{(a) DC current--voltage characteristics for one complete cycle showing bistable hysteresis at four different temperatures. (b) Variation of the critical current $I_{\rm c}$ and retrapping current $I_{\rm r}$ with bath temperature $T_{\rm b}$. The hysteresis crossover temperature is $T_{\rm h}$ = 6.86K. The red solid line is the fit to the $I_{\rm r}$ expression (see texts later).}
	\label{fig:2}
\end{figure}
The DC current--voltage characteristics (IVCs) of the device, shown in Fig. \ref{fig:2}(a), are strongly hysteretic with a high critical current ($I_{\rm c}$) and a retrapping current ($I_{\rm r}$) over a full (forward and reverse) cycle of bias-current sweep [$0, I$]. Above $I_{\rm r}$, the perfectly linear IVC with slope $\approx$ 30$\Omega$ confirms that the WL is fully ohmic with null superconducting phase coherence in the dissipative state. This is an (indirect) verification of the thermal origin of $I_{\rm r}$ and the bistability \cite{nikhil prl}, contrary to that in the RCSJ model \cite{squidbook}. The variations of $I_{\rm c}$ and $I_{\rm r}$ at different $T_{\rm b}$ are shown in Fig. \ref{fig:2}(b). The hysteresis crossover is found at temperature $T_{\rm h}$ = 6.86K, above which $I_{\rm r}\geq I_{\rm c}$ and the WL is nonhysteretic. Our focus is below $T_{\rm h}$.

\begin{figure*}
	\centering
	\includegraphics[width=2\columnwidth]{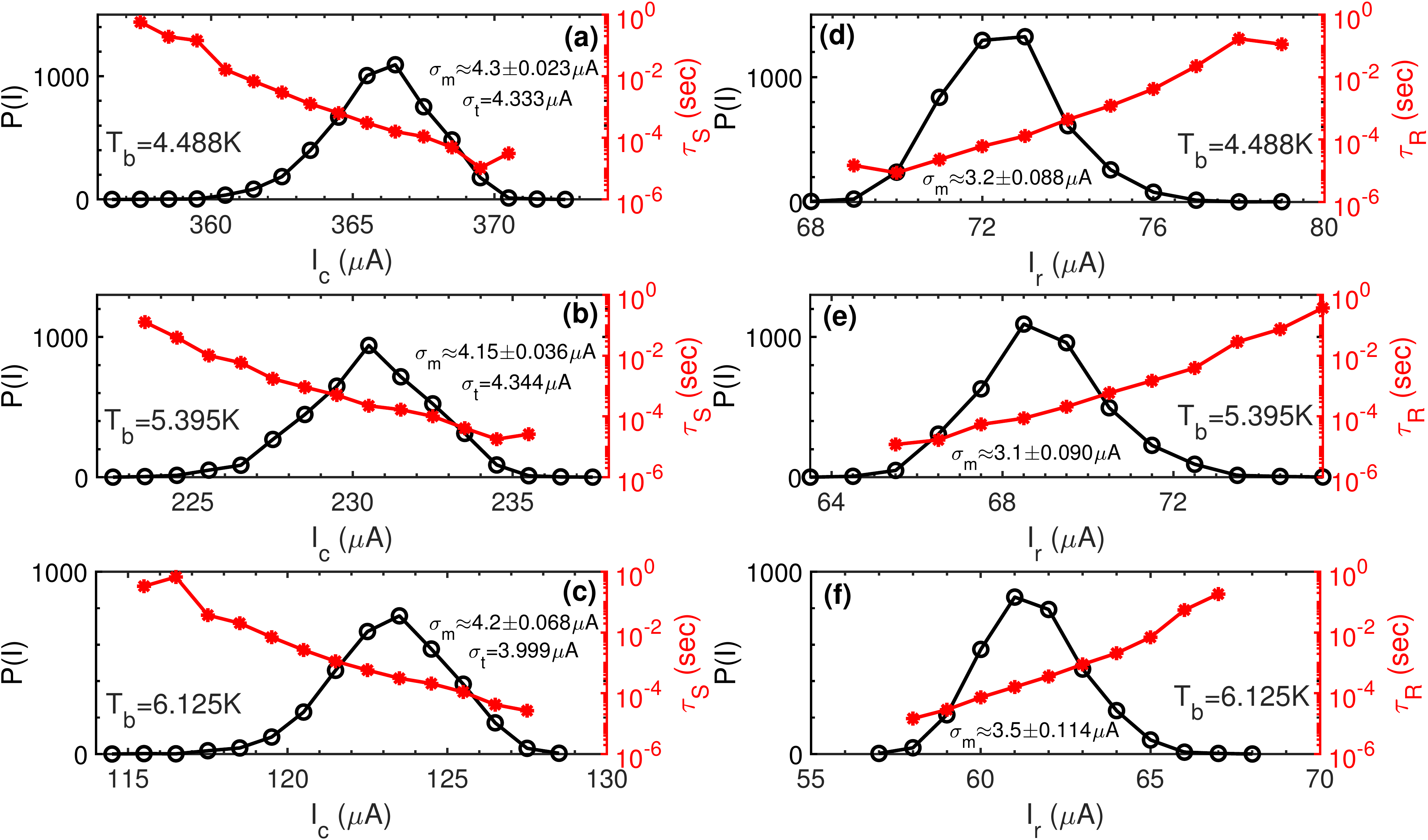}
	\caption{Histograms of the critical current $I_{\rm c}$ \big[(a-c)\big] and retrapping current $I_{\rm r}$ \big[(d-f)\big] at bath temperatures $T_{\rm b}$ = 4.488, 5.395 and 6.125K. The right \textit{y} axis represents the calculated lifetimes $\tau_S$ and $\tau_R$ of the superconducting and resistive states, respectively. $\sigma_{\rm m}$ and $\sigma_{\rm t}$ are the measured and calculated full width at half maximum of the distribution, respectively. The value of $I_{\rm c}$ at the peak is used in the calculation.}
	\label{fig:3}
\end{figure*}
\begin{figure*}
	\centering
	\includegraphics[width=2\columnwidth]{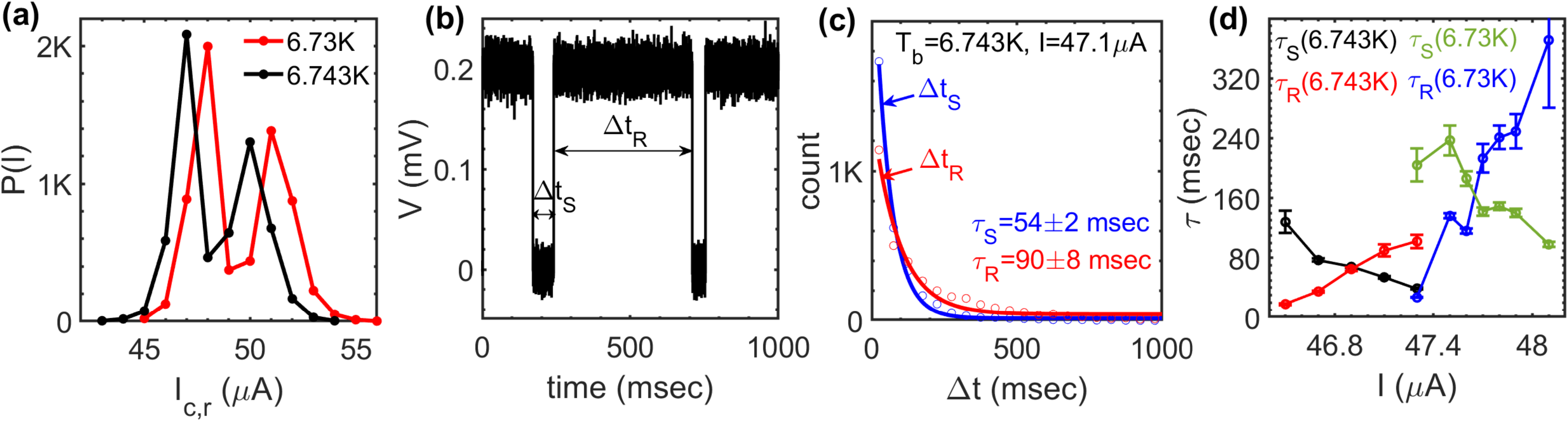}
	\caption{(a) Probability distribution curves showing overlap of $I_{\rm c,r}$ at two $T_{\rm b}$ points near $T_{\rm h}$. (b) A typical telegraphic switching over the time domain of $1000$ ms at a bias current, $I$ = 47.1$\mu$A and at $T_{\rm b}$ = 6.743K. Time spent ($\Delta t_{\rm S,R}$) at the two states is shown. (c) Distributions of the observed $\Delta t_{\rm S,R}$ are shown by symbols. Solid lines are fit to the exponential decay function. The decay constant gives the mean lifetime $\tau_{\rm S,R}$. (d) Variations of $\tau_{\rm S,R}$ with the bias current.}
	\label{fig:4}
\end{figure*}
The values of $I_{\rm c}$ and $I_{\rm r}$ are not unique in the hysteretic regime. For many full current cycles, we see different switching positions for both $I_{\rm c}$ and $I_{\rm r}$. To obtain a large number of these values, we bias the device with an AC sinusoidal current of frequency 5Hz and with peak amplitude sufficiently higher than the $I_{\rm c}$ value found in a DC IVC. Each cycle supplies a $I_{\rm c}$ and a $I_{\rm r}$. The histograms of $I_{\rm c,r}$ are plotted and the corresponding counts (or probability distributions $P(I)$) are shown in Fig. \ref{fig:3} for three bath temperatures $T_{\rm b}$.
We observe a small variation in the measured distribution widths $\sigma_{\rm m}$ within the limited bath temperature range. This is attributed to the dominant effect of the thermally activated phase-slip (TAPS) over the quantum phase-slip (QPS) at our (relatively high) working temperature. A single TAPS event at a bath temperature $T_{\rm b}$ gives the theoretical $I_{\rm c}$-distribution width $\sigma_{\rm t}$ according to $\sigma_{\rm t}\sim (k_{\rm B}T_{\rm b}/\Phi_{\rm 0})^{1/\beta}\times I_{\rm c}(T_{\rm b})^{1-1/\beta}$ \cite{pengli-prl}. Here, $k_{\rm B}$ is the Boltzmann constant, $\Phi_0$ is the flux quantum, and $\beta$ is an exponent\cite{Note} whose value depends on the junction dimension. The calculated $\sigma_{\rm t}$ is found to be close to the measured $\sigma_{\rm m}$ [see Fig. \ref{fig:3}(a)--\ref{fig:3}(c)], for an approximate $\beta=1.38$ in a longer WL limit. Note that the width in switching current histograms may depend on various factors such as the superconducting material, substrate, electronic noise; no universal temperature dependence of $\sigma$ is observed \cite{fluctuation-nanowire,bezryadin-2}.

$P(I)$ for the switching between $I$ and $I+dI$ is related to the lifetime $\tau$ at $I$ through the relation \cite{fulton}
\[P(I)=\frac{\tau}{dI/dt}\left(1-\int_{0}^{I}P(I)dI\right),\]
where $dI/dt$ is the ramp-rate of bias current. The estimated lifetimes for the zero-voltage superconducting state, ${\tau}_{\rm S}$ and for the nonzero-voltage resistive state, ${\tau}_{\rm R}$ from the respective $I_{\rm c,r}$ distributions are shown in Fig. \ref{fig:3} with right \textit{y} axis. The values span from $\sim10^{-5}$ s to $\sim1$ s.

As seen in Fig. \ref{fig:2}(b) that $I_{\rm c,r}$ come nearer with increased $T_{\rm b}$, we now set the working temperature very close to the hysteresis crossover temperature $T_{\rm h}$. Here, distinguishing between $I_{\rm c}$ and $I_{\rm r}$ in a cycle often becomes tedious; however, adjusting $T_{\rm b}$ to a suitable value, the obtained $P(I)$ for $I_{\rm c,r}$ are plotted [see Fig. \ref{fig:4}(a)] for two different $T_{\rm b}$. The distributions significantly overlap with each other. More interestingly, when a fixed (DC) bias current chosen from the overlapping region is fed, we observe random jumps between zero- and finite-voltage states showing a clear random telegraphic noise (RTN) in a bistable system \cite{souravaip,stochastic}. Figure \ref{fig:4}(b) shows a characteristic RTN at $T_{\rm b}=$ 6.743K and $I=47.1 \mu$A.

RTN in the time domain with two well-separated states explicitly reveals the states' lifetimes. From each two consecutive switching events, we extract $\Delta t_{\rm S}$ and $\Delta t_{\rm R}$ [see Fig. \ref{fig:4}(b)] which represent the time spent in the superconducting and resistive states, respectively. Fluctuation-driven random switching is a stochastic process; hence, $\Delta t_{\rm S}$ and $\Delta t_{\rm R}$ follow exponential distributions. For reliable statistical analysis of $\Delta t_{\rm S,R}$, we take more than 1000 switching events at each bias current. The count distributions for $\Delta t_{\rm S,R}$ along with their exponential fits are shown in Fig. \ref{fig:4}(c). The exponential decay constant is a measure of the mean lifetime ${\tau}_{\rm S,R}$ for the respective (metastable) state. Thus-measured ${\tau}_{\rm S,R}$ and the variation with bias current at two different bath temperatures are shown in Fig. \ref{fig:4}(d). As the bias current is increased near the transition, the device favors more the resistive state; hence, $\tau_{\rm R} (\tau_{\rm S})$ increases (decreases) and they cross at a point. Limited by the resolution of our electronics, RTN provides the (minimum) timescale up to the order of $\sim10^{-2}$ s. We show the RTN in the hysteretic regime here, while the same in a nonhysteretic regime has been reported as a technique to increase micro-SQUID sensitivity \cite{sagarnano}.

\section{Thermal bistability model}
Let us first briefly discuss the two states of an ideal JJ based on the RCSJ model. This model explains the transport properties of the junction with a so-called tilted washboard potential which has the form: $E=\frac{\hbar I_{\rm c}}{2e}(\cos\varphi+I/I_{\rm c})$ \cite{tinkham book}. The superconducting phase $\varphi$ is analogous to a point-mass. Superconducting state with (constant) $\varphi$ of the JJ is depicted by the point-mass trapped in a potential minimum; thus, $\varphi$ is static when $I<I_{\rm c}$. With increasing $I$, $E$ becomes further tilted, and for $I\geq I_{\rm c}$, the minima no longer exist. Thus, at $I_{\rm c}$, $\varphi$ (the mass) starts rolling down the potential, giving rise to a voltage ($\equiv d\varphi/dt$) in the junction according to the AC Josephson relation. In a physical JJ, due to thermal ($k_{\rm B}T$) and/or quantum fluctuations, $\varphi$ can cross the height of the potential-barrier given by $\Delta E_{\rm S}=\frac{\hbar I_{\rm c}}{2e}(1-I/I_{\rm c})^{3/2}$, at a $I$ below the intrinsic $I_{\rm c}$. This phase-slip phenomenon leads to a distribution of $I_{\rm c}$ for the transition from the zero-voltage superconducting state to the finite-voltage state. The lifetime $\tau_{\rm S}$ is related to the barrier height via the Arrhenius equation of activation energy: $\frac{1}{\tau_{\rm S}}=\omega\exp(-\frac{\Delta E_{\rm S}}{k_{\rm B}T})$, where $\omega$ is the attempt frequency due to phase-slip \cite{phaseslip,nayana}. Using the RCSJ equations, our observed results of $\tau_{\rm S}$ are analyzed in Appendix-B.


During the reverse current cycle, the dynamic phase becomes static again at a retrapping current depending on the impedance (damping) of the washboard potential, which is determined by the junction capacitance. Using the RCSJ model, the switching statistics were studied \cite{fulton,prl2}, and telegraphic noise near crossover temperature was simulated \cite{rtnsimulation}. The switching between metastable states was alternatively explained by the time-dependent Ginzburg--Landau theory \cite{tdgl}. Thermal fluctuation effect on the steady-state of a JJ was also modeled within a nonequilibrium approach \cite{benjacob2}.

A Josephson WL having a larger $I_{\rm c}$ than that of a tunnel-barrier type conventional JJ, exhibits Joule heating due to $I_{\rm c}^2R_{\rm WL}$. The latter gives rise to a temperature increase at the junction. The RCSJ model does not take such heat dissipation into account in the finite-voltage state and, hence, is not plausible for explaining $I_{\rm r}$ and the bistability in a WL or nanowire. Moreover, the geometric capacitance, which is central in the RCSJ model, is negligible for such a JJ. Thus, a RSJ model is appropriate for a moderately heated WL with its raised temperature $T$ ($T_{\rm b}<T<T_{\rm c}$), where the Josephson coupling \cite{sourav1,shapiro} can sustain in the dissipative state. However, for much larger $I_{\rm c}$ values at low bath temperatures ($T_{\rm b}<T_{\rm h}$), WL temperature remains $T>T_{\rm c}$ due to enormous heating in the dissipative state for $I\geq I_{\rm r}$. Therefore, the Josephson coupling is fully lost and $\varphi$ is irrelevant. One, thus, needs to solve static heat balance equations \cite{scocpol,tinkham nano wire} with the thermal parameters of the device.
\begin{figure}
	\centering
	\includegraphics[width=\columnwidth]{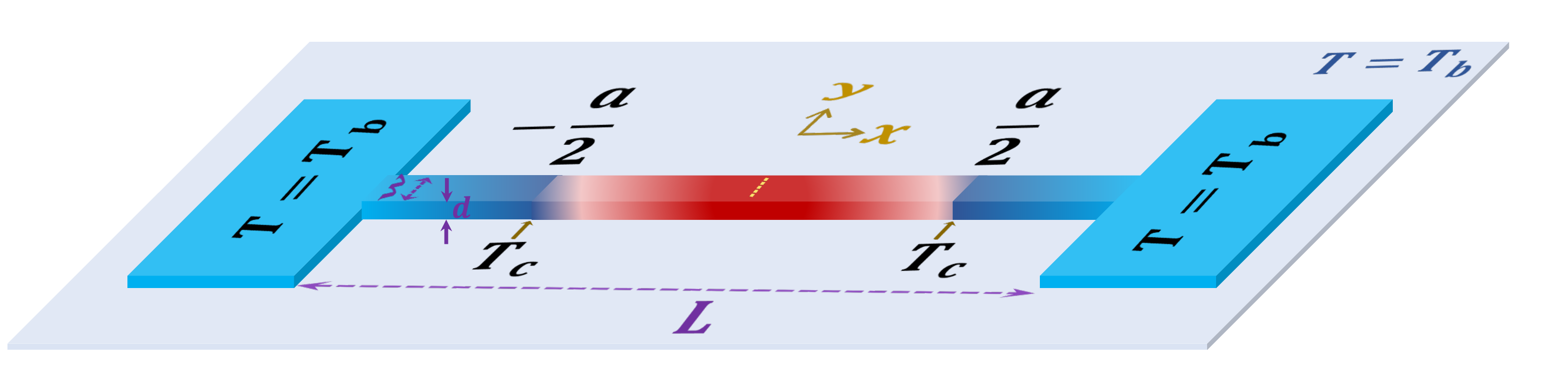}
	\caption{Schematic diagram showing normal (N, shown by red) and superconductor (S, shown by blue) regime along the lead in the dissipative state. $L$, $w$, and $d$ are the length, width, and thickness of the lead, respectively. The NS interface is formed at $x=\pm a/2$.}
	\label{fig:5}
\end{figure}

\begin{figure}
	\centering
	\includegraphics[width=\columnwidth]{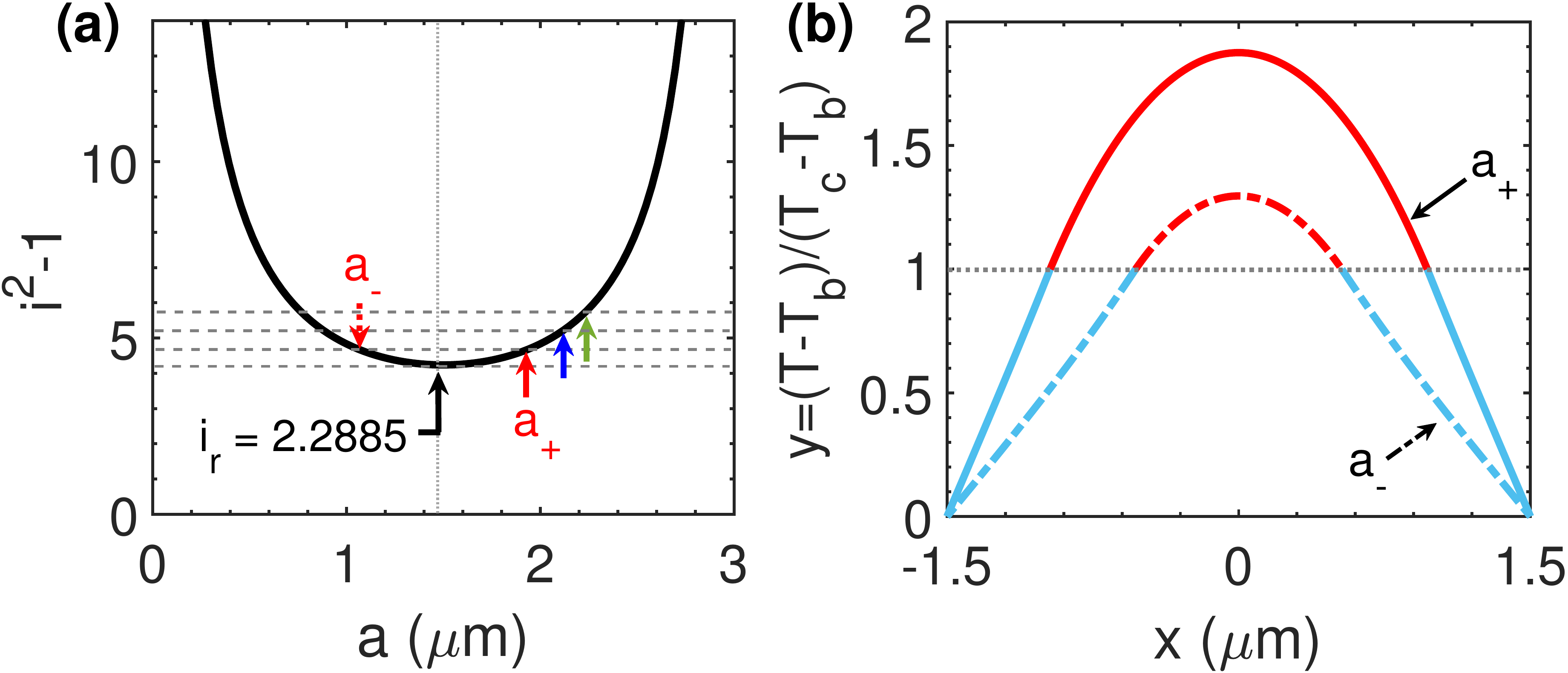}
	\caption{(a) ($i^2-1$) vs $a$ plot for the length scale of 3$\mu$m, showing a minimum at $a=L/2$ with $i=i_{\rm r}=2.2885$. Four dashed horizontal lines corresponding to $i=i_{\rm r}, 2.3843, 2.4951, 2.605$ are drawn. Each cut gives two solutions ($a_{+},a_{-}$) for $a$, pointed by arrows. (b) Temperature profiles along the lead for the two solutions of $a$ marked by the red arrows in (a). Solid (dashed) line is for stable (unstable) NS interface at $a_+$ ($a_-$). $y=1$ implies the NS interface.}
	\label{fig:6}
\end{figure}

We model our device as a superconducting lead connected between two reservoirs ($T_{\rm b}$). Above $I_{\rm r}$, a part of the lead is normal (N) with temperature $T>T_{\rm c}$ and rest is superconductor (S) with $T<T_{\rm c}$. Figure \ref{fig:5} shows the schematic with NS interface at $x=\pm a/2$ for a bias $I>I_{\rm r}$. The heat balance equations for N and S regimes in the steady-state are given by 
\begin{align}
-K\frac{d^{2}T}{dx^2}+\frac{\alpha}{d}(T-T_b)&=\frac{I^2}{(wd)^2}\rho \quad \quad \text{for}\quad |x|\leq \frac{a}{2}, \label{eq:1}\\
-K\frac{d^{2}T}{dx^2}+\frac{\alpha}{d}(T-T_b)&=0 \quad \quad \quad \text{for}\quad |x|\geq \frac{a}{2}. \label{eq:2}
\end{align}
Here, $K$ is the temperature-independent thermal conductivity of both the N and S regime, $\alpha$ is the heat transfer coefficient, and $\rho$ is the resistivity. In terms of nondimensional temperature ($y$) given by $y=\frac{T-T_{\rm b}}{T_{\rm c}-T_{\rm b}}$ and bias current ($i$) given by $i^2=\frac{I^2}{I_0^2}$, Eqs. (\ref{eq:1}) and (\ref{eq:2}) can be written as
\begin{align}
\eta^2\frac{d^{2}y_{\rm n}}{dx^2}-y_{\rm n}+i^2&=0 \quad \quad \text{for}\quad |x|\leq \frac{a}{2}, \label{eq:3}\\
\eta^2\frac{d^{2}y_{\rm s}}{dx^2}-y_{\rm s}&=0 \quad \quad \text{for}\quad |x|\geq \frac{a}{2}, \label{eq:4}
\end{align}
where $\eta=\sqrt{\frac{Kd}{\alpha}}$, called the thermal healing length \cite{scocpol}. $I_0$ is given by $I_0^2=\frac{w^2d\alpha(T_c-T_b)}{\rho}$. The solutions of these differential equations are
\begin{align}
y_{\rm n}(x)&=i^2+C_1\exp\Big(\frac{x}{\eta}\Big)+C_2\exp\Big(\frac{-x}{\eta}\Big)  \quad \text{for}\quad |x|\leq \frac{a}{2}, \label{eq:5}\\
y_{\rm s}(x)&=C_3\exp\Big(\frac{x}{\eta}\Big)+C_4\exp\Big(\frac{-x}{\eta}\Big) \quad \quad \text{for}\quad |x|\geq \frac{a}{2}, \label{eq:6}
\end{align}
and $C_{(1-4)}$ are the constants. Using four boundary conditions: $y_{\rm s}(\frac{L}{2})=0$, $y_{\rm n}(\frac{a}{2})=y_{\rm s}(\frac{a}{2})$, $y_{\rm n}(\frac{a}{2})=y_{\rm s}(\frac{a}{2})=1$ and $\frac{dy_{\rm n}}{dx}=0$ at $x=0$ (due to symmetry), we obtain the steady state temperature profiles in N and S regimes,
\begin{align}
y_{\rm n}(x)&=i^2-(i^2-1)\frac{\exp\big(\frac{x}{\eta}\big)+\exp\big(\frac{-x}{\eta}\big)}{\exp\big(\frac{a}{2\eta}\big)+\exp\big(\frac{-a}{2\eta}\big)} \quad \quad \text{for}\quad |x|\leq \frac{a}{2}, \label{eq:7} \\
y_{\rm s}(x)&=\frac{\exp\big(\frac{-x}{\eta}\big)-\exp\big(\frac{x}{\eta}-\frac{L}{\eta}\big)}{\exp\big(\frac{-a}{2\eta}\big)-\exp\big(\frac{a}{2\eta}-\frac{L}{\eta}\big)}\quad \quad \text{for}\quad |x|\geq \frac{a}{2}. \label{eq:8}
\end{align}
We use the fact that the above two solutions must satisfy the following condition at NS interface ($x=\frac{a}{2}$): $\frac{dy_{\rm n}}{dx}=\frac{dy_{\rm s}}{dx}$. This would lead to an equation for $i$ as
\begin{equation}
(i^2-1)=\coth\Big(\frac{-L}{2\eta}+\frac{a}{2\eta}\Big)\times\coth\Big(\frac{-a}{2\eta}\Big).
\label{eq:9}
\end{equation}
Equation (\ref{eq:9}) shows a minimum at $a=\frac{L}{2}$ in $a$ vs $(i^2-1)$ plot [see Fig. \ref{fig:6}(a)]. Putting $a=\frac{L}{2}$ in Eq. \ref{eq:9}, we get the expression for retrapping current, $i_{\rm r}=\sqrt{2}\times\sqrt{\frac{\exp\big(-L/2\eta\big)+\exp\big(L/2\eta\big)}{\exp\big(-L/2\eta\big)+\exp\big(L/2\eta\big)-2}}$. This is the minimum bias current required for an NS interface to stabilize. Below $i<i_{\rm r}$, the WL and whole lead are superconducting.

In our device, the NS interface extends beyond the WL up to the narrow leads, as estimated from the resistance of 30$\Omega$ above $I_{\rm r}$. With the WLs' contribution of 17.9$\Omega$ and the estimated $R_{\rm sh}$, the normal portion of the device above $I_{\rm r}$ is roughly marked by AB [see Fig. \ref{fig:1}(b)]. Taking the approximate length of $L\approx3$$\mu$m, $w=300$nm, and $d=20$nm, we fit the observed $I_{\rm r} (T_{\rm b})$ to the above $i_{\rm r}$ expression [see Fig. \ref{fig:2}(b)]. The obtained fitting parameter is $\alpha\approx2$ W/cm$^2$K. We used $K=\mathcal{L}T_{\rm c}/\rho$, where the Lorentz number is $\mathcal{L}=2.44\times10^{-8}$ W$\Omega$/K$^2$, $T_{\rm c}=7.78$K (the second transition for WL and nearest leads, shown in Fig. \ref{fig:1}(c)) and $\rho=9.5$ $\mu\Omega$.cm. For these device parameters, $i_{\rm r}=2.2885$ and $I_0=19.94$$\mu$A (at $T_{\rm b}=6.73$K).

\begin{figure}
		\centering
	\includegraphics[width=\columnwidth]{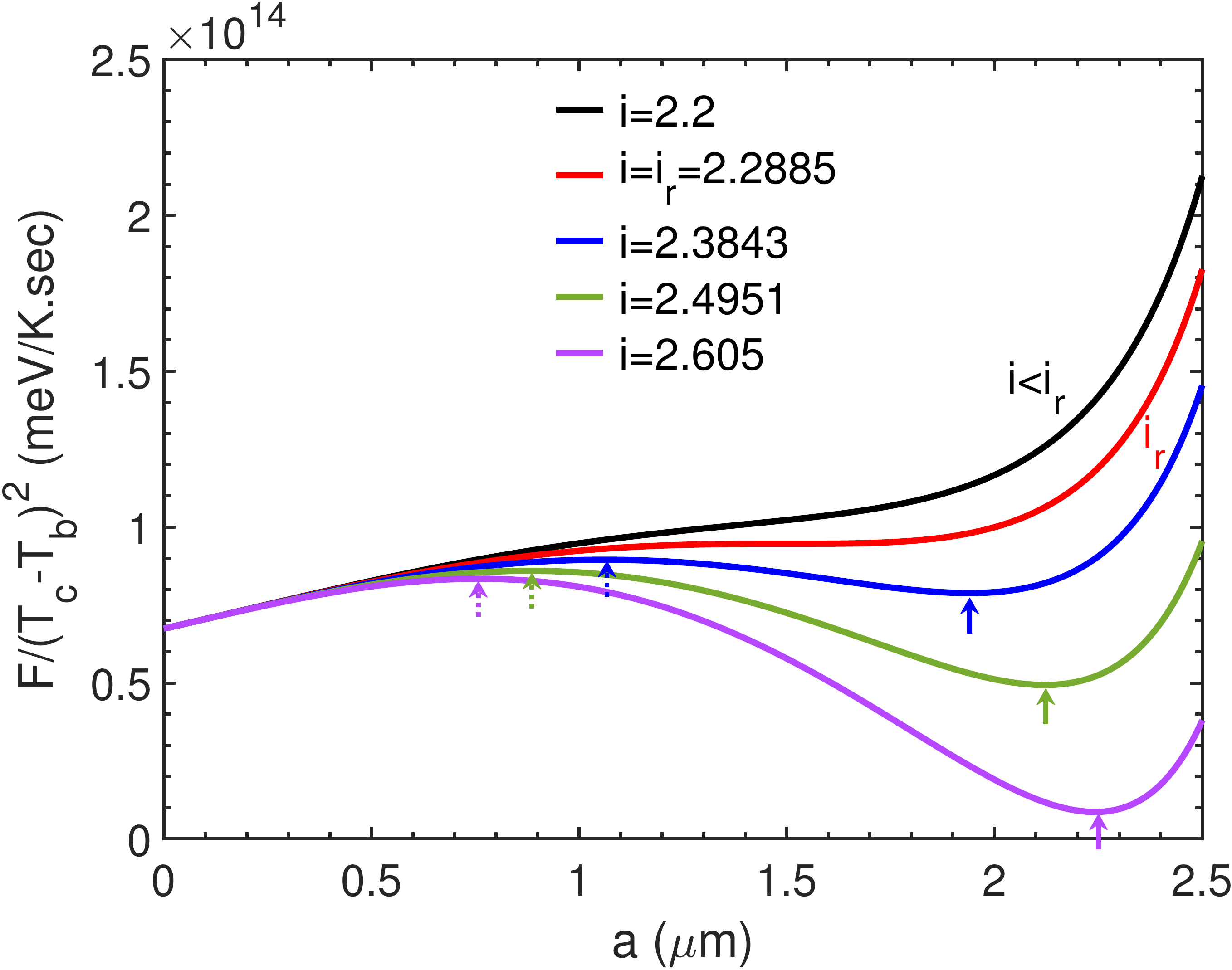}
	\caption{$F/(T_{\rm c}-T_{\rm b})^2$ as a function of $a$ showing double-well feature with a minimum (indicated by the solid arrow) and maximum (indicated by the dashed arrow) for $i>i_{\rm r}$. Minimum of $F$ vanishes below $i_{\rm r}$.}
	\label{fig:7}
\end{figure}

For $i>i_{\rm r}$, Eq. \ref{eq:9} has two solutions for $a$; one is for $a>L/2$, say, $a_+$, and the other for $a<L/2$, say, $a_-$ [see Fig. \ref{fig:6}(a)]. However, for $a=a_-$, the NS interface is unphysical because with increasing $i$, the normal regime would decrease. Thus, the stable NS interface occurs at $a=a_+$ for $I>I_{\rm r}$. With the obtained $a$, the temperature profile is given by Eqs. (\ref{eq:7},\ref{eq:8}). Figure \ref{fig:6}(b) displays the two temperature profiles over the lead for the two (stable and unstable) $a$ values at $i=2.3843$.
 
The stability of the NS interface is subjected to thermal fluctuations. To obtain stability, we consider the heat balance at any instantaneous time $t$. Therefore, the time-dependent thermal equations are given by
\begin{align}
C\frac{\partial T}{\partial t}&=\frac{I^2}{(wd)^2}\rho+K\frac{d^{2}T}{dx^2}-\frac{\alpha}{d}(T-T_{\rm b}) \quad \quad \text{for}\quad |x|\leq \frac{a}{2}, \label{eq:10}\\
C\frac{\partial T}{\partial t}&=K\frac{d^{2}T}{dx^2}-\frac{\alpha}{d}(T-T_{\rm b}) \quad \quad \text{for}\quad |x|\geq \frac{a}{2},\label{eq:11}
\end{align}
which, in steady states, would converge to Eqs. (\ref{eq:1}) and (\ref{eq:2}). $C$ is the specific heat of the material.
Let us now define the following integral:
\begin{eqnarray}\nonumber
F=\int_{-\frac{L}{2}}^{\frac{L}{2}}\Big[\frac{1}{2}KA\Big(\frac{\partial T}{\partial x}\Big)^2+\frac{\alpha w}{2}(T-T_{\rm b})^2\\
-\frac{I^2 \rho}{A}(T-T_{\rm c})\Theta(T-T_{\rm c})\Big]dx
\label{eq:12}
\end{eqnarray}
such that $C\frac{\partial T}{\partial t}=-\frac{\partial F}{\partial T}$. Integrand of Eq. \ref{eq:12} is a function of ($T,\frac{\partial T}{\partial x},x$) and it follows the Euler--Lagrange equation. Note that $\Theta(T-T_{\rm c})$ is $1$ for $|x|\leq \frac{a}{2}$ and $0$ for $|x|\geq \frac{a}{2}$. Applying variational principle \cite{variational-1,variational-2,variational-interface} for small variation in temperature $T\Rightarrow T+\delta T$ in stationary states, $\delta F=0$ gives rise to Eqs. (\ref{eq:1}) and (\ref{eq:2}). Therefore, $F$ represents the physical quantity of the system, which is an extremum in its stable and unstable state.

We now simplify the quantity $F$ for the temperature $T(x)$ along the lead. $F$ is written as follows:
\begin{widetext}
\begin{equation}
F=2\Big[\int_{0}^{\frac{a}{2}}\Big\{\frac{1}{2}KA\Big(\frac{\partial T}{\partial x}\Big)^2+\frac{\alpha w}{2}(T-T_{\rm b})^2-\frac{I^2 \rho}{A}(T-T_{\rm c})\Big\}dx
+\int_{\frac{a}{2}}^{\frac{L}{2}}\Big\{\frac{1}{2}KA\Big(\frac{\partial T}{\partial x}\Big)^2+\frac{\alpha w}{2}(T-T_{\rm b})^2\Big\}dx\Big].
\label{eq:13}
\end{equation}
In terms of dimensionless temperature $y$, Eq. \ref{eq:13} is given by
\begin{equation}
F=2\alpha w(T_{\rm c}-T_{\rm b})^2\Big[\int_{0}^{\frac{a}{2}}\Big\{\frac{\eta^2}{2}\Big(\frac{\partial y_{\rm n}}{\partial x}\Big)^2+\frac{1}{2}y_{\rm n}^2-i^2(y_{\rm n}-1)\Big\}dx
+\int_{\frac{a}{2}}^{\frac{L}{2}}\Big\{\frac{\eta^2}{2}\Big(\frac{\partial y_{\rm s}}{\partial x}\Big)^2+\frac{1}{2}y_{\rm s}^2\Big\}dx\Big].
\end{equation}
Using Eqs. (\ref{eq:7}) and (\ref{eq:8}) for $y_{\rm n}(x)$ and $y_{\rm s}(x)$, respectively, we obtain the expression for $F$:

\begin{equation}
F=2\alpha w(T_{\rm c}-T_{\rm b})^2\Big[-\frac{1}{2}\eta \coth\Big(\frac{a-L}{2\eta}\Big)+\frac{1}{4}\Big\{-ai^2(-2+i^2)+2(-1+i^2)^2\eta \tanh\Big(\frac{a}{2\eta}\Big)\Big\}\Big].
\end{equation}
\end{widetext}
The variation of $F/(T_{\rm c}-T_{\rm b})^2$ with $a$ is plotted in Fig. \ref{fig:7} for different $i$. Below $i<i_{\rm r}$, no minimum exists in $F$; above $i>i_{\rm r}$, a minimum and an adjacent maximum appear. This minimum (maximum) corresponds to the stable (unstable) NS interface discussed before.
\begin{figure}
	\centering
	\includegraphics[width=\columnwidth]{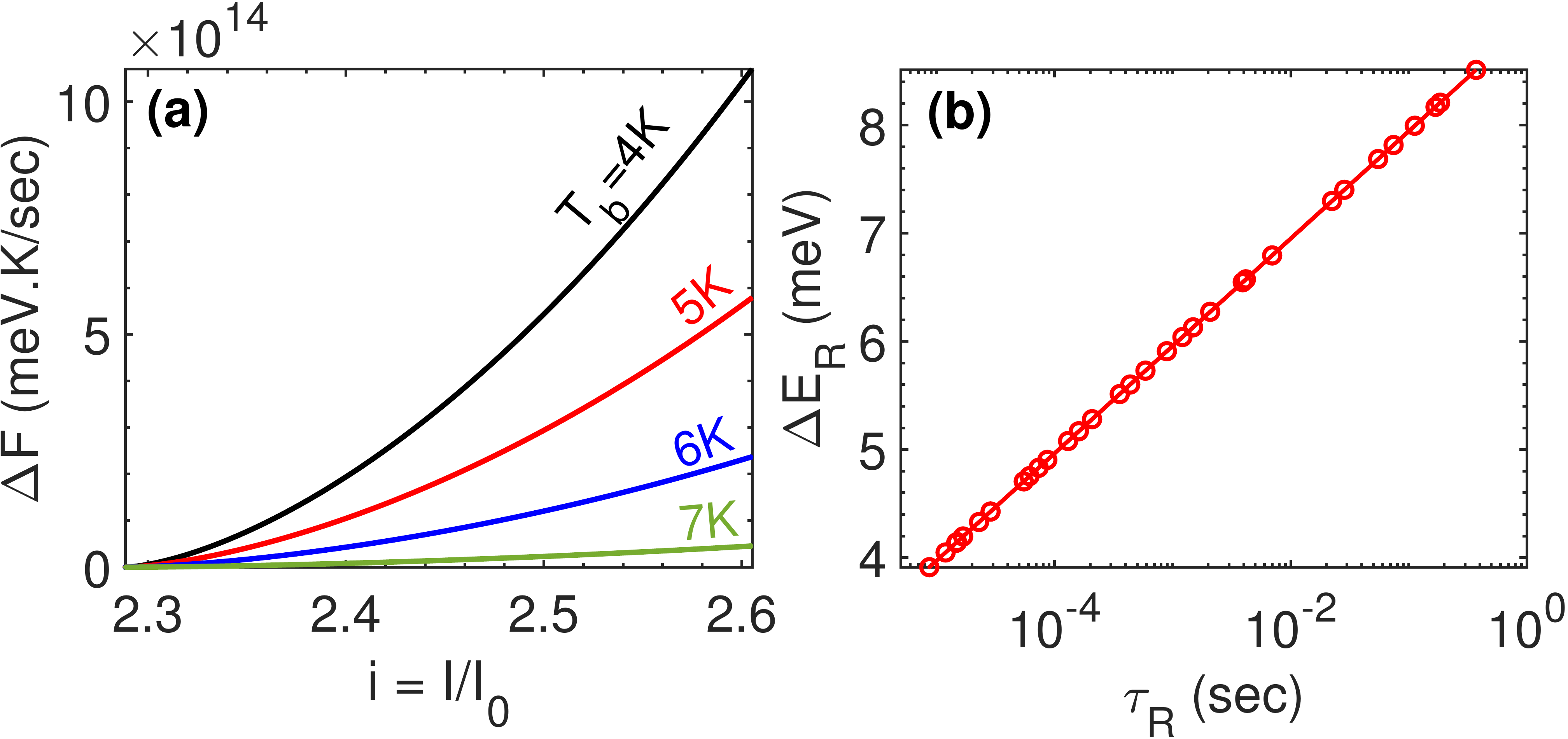}
	\caption{(a) Variation of $\Delta F$ with $i$ ($>i_{\rm r}$) for different $T_{\rm b}$. (b) Typical barrier height (activation energy) $\Delta E_{\rm R}$ in the resistive state calculated from the experimentally obtained $\tau_{\rm R}$.}
	\label{fig:8}
\end{figure}
\section{Discussion}
In the finite-voltage resistive (metastable) state above the retrapping current, the device belongs to a $F$ minimum. At finite temperatures, the $k_{\rm B}T$ thermal fluctuations can be enough to cross the nearest maximum leading the device to jump into the fully superconducting state. In other words, the thermal fluctuations cause the stable NS interface at $a_+$ to fluctuate; once it reaches the unstable position at $a_-$, it will collapse. This is demonstrated by incorporating a heat pulse in Eqs. (\ref{eq:10}) and (\ref{eq:11}) [see Appendix-C]. Thus, a distribution in $I_{\rm r}$ is observed. When $I_{\rm c}$ and $I_{\rm r}$ are close, the junction exhibits $\sim1/\tau_{\rm ps}$ rate of phase-slip each depositing $I\Phi_{\rm 0}$ heat in the superconducting state, which pushes it toward the resistive state again. Thus, an RTN is observed.

With increasing bias current, the difference $\Delta F$ between the minimum and maximum increases [see Fig. \ref{fig:8}(a)]; so does the lifetime ($\tau$) of the resistive state as observed in the experiment. To quantify $\tau$ against fluctuations, similar to the thermal activation (Arrhenius) analysis, the right potential (or free energy) expression is desired. However, one could notice that though $F$ is minimum when the system is stable, it does not provide any such ``energy''; rather it is of the dimension of (Energy$\times$Temperature/Time). The latter appears to be a quantity not usually seen in any physical system. This unveiled quantity $F$ for a superconducting device with dissipation, which represents the stability of its states, is intriguing. However, not being potential energy, the incorporation of attempt frequency and, hence, a quantitative estimation of $\tau$ from $F$ is not clear to us; although the model intuitively captures the spread in $I_{\rm r}$ and lifetime variation with the bias. Moreover, in a dissipative system, the escape rate from a metastable minimum is significantly influenced by the dissipation \cite{dissipation,dissipation-1}. We, thus, believe that finding lifetime likely be nontrivial for a heat-dissipative WL system, which necessitates advanced theoretical work.

Nevertheless, with our experimental data of $\tau_{\rm R}$, we estimate the associated energy scale $\Delta E_{\rm R}$ from the equation $\frac{1}{\tau_{\rm R}}=\frac{1}{\tau_{\rm th}}\exp(-\frac{\Delta E_{\rm R}}{k_{\rm B}T})$. Here, $\tau_{\rm th}$ is thermal time, which is of the order of $\sim 10^{-9}$ s \cite{anjanjap,scocpol}. Figure \ref{fig:8}(b) shows the linear variation of $\Delta E_{\rm R}$ with $\log\tau_{\rm R}$. Typical value of $\Delta E_{\rm R}$ is few meV.
 
The switching behavior at the critical current $I_{\rm c}$ enables a superconducting WL or nanowire to detect single photons \cite{spd}. To detect such tiny energy, a current biased WL is set at a DC bias current just below the $I_{\rm c}$. Upon absorbing the incident photon, the Cooper pair breaks into the normal electron making the WL resistive and producing a voltage spike. However, close to $I_{\rm c}$, ubiquitous (intrinsic and extrinsic) fluctuations due to thermal or quantum phase-slip, bias noise, substrate, etc. can also trigger the same \cite{phaseslip,fluctuation-sc,fluctuation-nanowire,biasnoise}. Therefore, it has usually been a challenge to distinguish the effect of the noise and to get rid of the false counts \cite{nanowire-challenge}.

Moreover, in the finite-voltage state, a local hot-spot with an NS interface is generated. Its dynamics with thermal and electrical parameters in the device precisely dictate the $I_{\rm r}(<I_{\rm c})$, which can also be thought of as a reduced critical current due to raised local temperature. The hot-spot characteristics are a crucial part of a WL employed as a detector. The former requires cooling down via a thermal relaxation process to come back to the initial superconducting state and subsequently detect the next signal \cite{hotspot-relax,hotspot,hotspot-1,hotspot-relax-1}. This determines the resetting time of the device, the resolution (frequency) of the detection and, thus, the efficiency \cite{spd}. While a perfect heat evacuation can eliminate the hysteresis completely \cite{anjanjap,sourav1}, a moderate evacuation increasing $I_{\rm r}$ apparently sounds beneficial. However, the inescapable RTN effect when $I_{\rm c,r}$ are close by must be considered. Therefore, our results will be instrumental for a suitable choice of the bias current and temperature and for an appropriate analysis in single-photon detection using WL or nanowire. The bistable model providing heat dynamics with material parameters in the resistive state will guide to thermally optimize such devices.

\section{Conclusion}
We studied the fluctuation-induced statistics of critical ($I_{\rm c}$) and retrapping ($I_{\rm r}$) currents of strongly hysteretic Josephson WL made of niobium. The estimated lifetimes ($\tau$) of the two metastable states, \mbox{i.e.,} zero-voltage superconducting state and nonzero-voltage resistive state, from the $I_{\rm c,r}$ distributions dictate the states' stability. Observed random telegraphic switching between the two states in a certain temperature and bias current regime is the direct experimental evidence of the decay process due to fluctuations in a WL two-state system. $\tau$ was also measured from the telegraphic data. We described a simple thermal model introducing a normal-superconductor interface and bistable characteristics. Putting forward a new analysis, we illustrated the transition between the states and the spread in $I_{\rm r}$. In terms of the applications of Josephson WL-based devices in the two-state systems, single-photon detectors, and resonators, our study could provide insights to develop more understanding for device improvement. To end, our model opens an interesting theoretical aspect for the future.

\section*{acknowledgments}
S.B. and A.K.G. acknowledge Herv\'e Courtois and Clemens B Winkelmann for a visiting position at Institut N\'eel, and for discussions. We acknowledge Jayanta K. Bhattacharjee and Satya Majumdar for a discussion. We appreciate the technical help of Thierry Crozes in making the devices using platform nanofab at Institut N\'eel, Grenoble. We acknowledge financial support from the Council of Scientific and Industrial Research (CSIR) and the Department of Science and Technology (DST) of the Government of India.
\section*{Author Declarations}
\subsection*{Conflict of interests}
The authors declare that there are no conflicts of interests.
\subsection*{Author contributions}
S.B. and A.K.G. contributed to device fabrication, measurements, analysis, and theoretical model. P.W. contributed to the theoretical model. S.B. wrote the manuscript with input from other authors.
\section*{Data Availability}
The data that support the findings of this study are available from the corresponding author upon reasonable request.

\section*{Appendix A: Calculation of sheet resistance}
\begin{figure}[h!]
	\centering
	\includegraphics[width=\columnwidth]{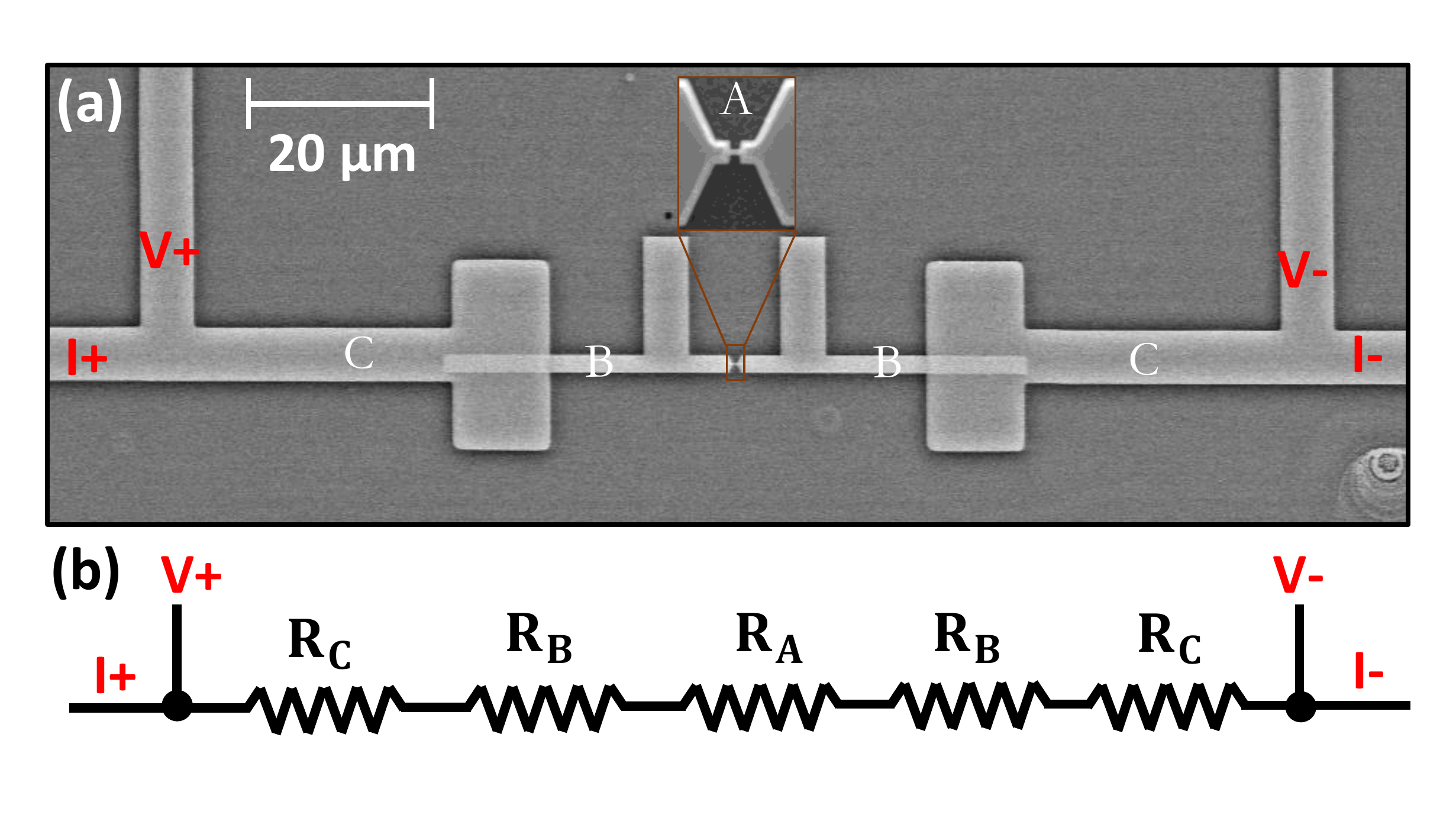}
	\caption{(a) Larger scale SEM image of the WL device with narrow, wide leads and current, voltage leads. Inset is the zoomed WL section. (b) Equivalent resistor circuit model.}
	\label{fig:9}
\end{figure}

Figure \ref{fig:9}(a) shows the SEM image of the whole device pattern between current and voltage pads. The length ($L$) and the width ($w$) of the WL are 160 and 40nm, respectively. The adjacent narrow leads, which are followed by the kinks [see inset of Fig. \ref{fig:9}(a)], are of dimension: $L=200$nm and $w=300$nm. The current and voltage leads are connected via two different-sized leads named B and C on either side. The dimensions ($L\times w$) of B and C leads are ($19\mu$m$\times 2\mu$m) and ($39\mu$m$\times 5\mu$m).

The structure can be modeled as a network of several series resistors [see Fig. \ref{fig:9}(b)] corresponding to the normal state resistance of various sections (A, B, C) just above the critical temperature $T_{\rm c}$. Thus, the total resistance ($R$) between the voltage leads is given by $R = R_{\rm A} + 2R_{\rm B} + 2R_{\rm C}$. Here, $R_{\rm A}$, $R_{\rm B}$, and $R_{\rm C}$ are the resistance of section A (the WL and nearest leads), lead B, and lead C, respectively. In terms of the resistivity per unit film thickness, which is called sheet resistance $R_{\rm sh}$, $R_{\rm A}=R_{\rm WL}+2\times\frac{200}{300}R_{\rm sh}$, $R_{\rm B}=\frac{19}{2}R_{\rm sh}$, and $R_{\rm C}=\frac{39}{5}R_{\rm sh}$. Using the obtained $R=161\Omega$ above the onset $T_{\rm c}$, we have $161=R_{\rm WL}+35.9R_{\rm sh}$. Considering dominant contributions from the leads and ignoring $R_{\rm WL}$, the estimated sheet resistance is $4.48\Omega$. The resistivity is $(R_{\rm sh}\times d)\sim 9.5$ $\mu\Omega$.cm.

\section*{Appendix B: Analysis using RCSJ model}
\begin{figure}[h!]
	\centering
	\includegraphics[width=0.85\columnwidth]{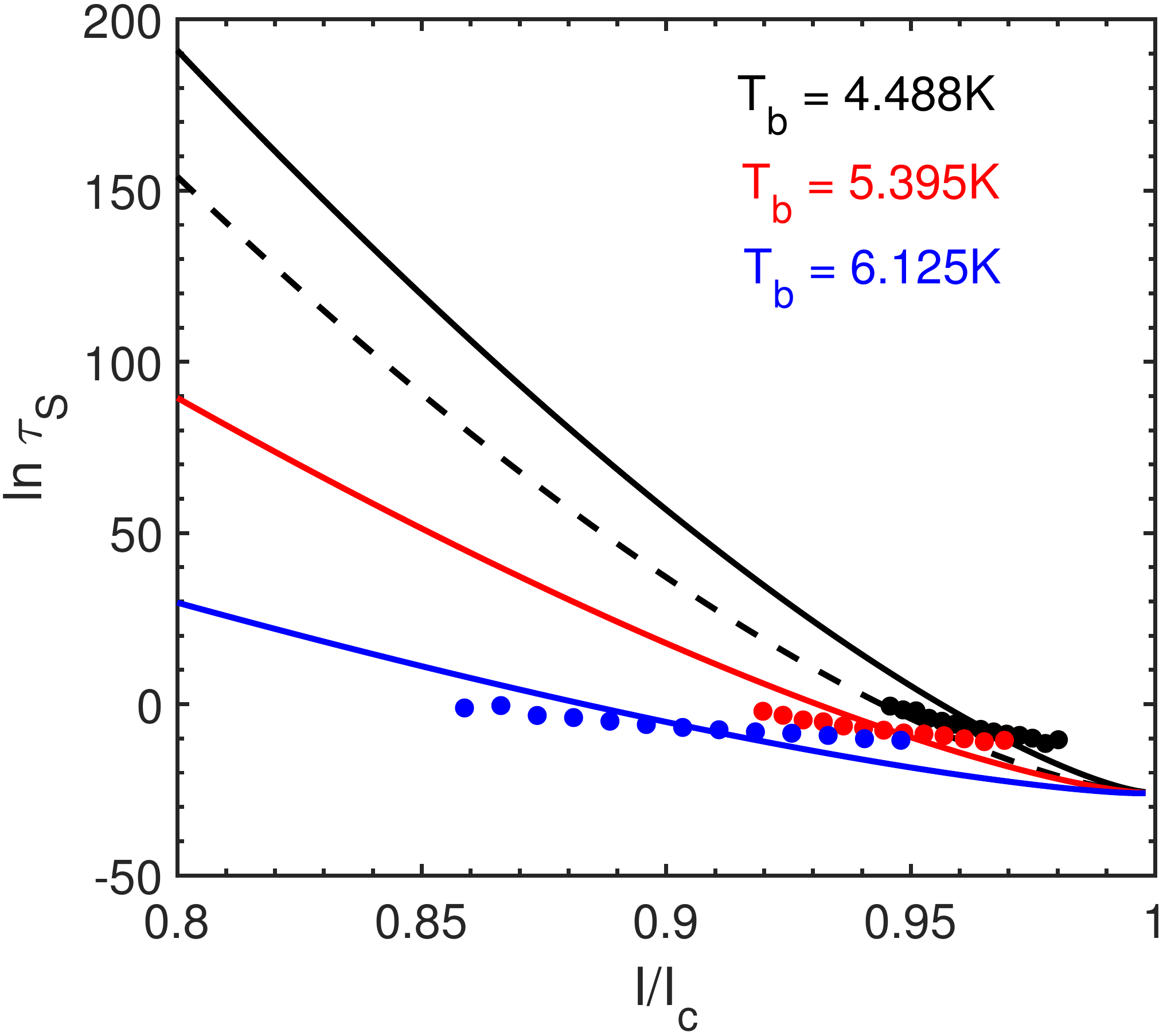}
	\caption{Variation of superconducting states' lifetime $\tau_{\rm S}$ with bias current. Dotted lines are the measured data. Solid lines are the approximate fit to the RCSJ model with $\beta=1.38$. The dashed line is for $\beta=3/2$, ideal JJ.}
	\label{fig:9.1}
\end{figure}
According to the RCSJ model, $\frac{1}{\tau_{\rm S}}=\omega\exp(-\frac{\Delta E_{\rm S}}{k_{\rm B}T})$. The attempt frequency is given by $\omega=\omega_0(1-I/I_{\rm c})^{\gamma}$. As discussed in the main text, $\Delta E_{\rm S}=\frac{\hbar I_{\rm c}}{2e}(1-I/I_{\rm c})^{\beta}$. $\beta$ ($\gamma$) varies between 3/2 (1/4) and 1 (0) as the WL length varies from the shortest to the longest limit compared to the superconducting coherence length \cite{likharev,tinkham book,moge-nanowire}. Using the $\tau_{\rm S}$ expression with $\beta=1.38$ and $\gamma=0.25$, we fit the obtained results from our measurements (see Fig. \ref{fig:9.1}). We stress that $I_{\rm c}$ in the equation is the intrinsic temperature-independent critical current. For each bath temperature $T_{\rm b}$, we use the $I_{\rm c}$ as the adjusting parameter. The obtained $I_{\rm c}$ values come out to be 378, 243, and 135$\mu$A for $T_{\rm b}$= 4.488, 5.395, and 6.125K. Note that the intrinsic $I_{\rm c}$ expectedly is always larger than the average $I_{\rm c}$ in a distribution. Also, note the difference in slopes of theoretical and experimental data. This could be due to the Joule heating effect, which is ignored in the RCSJ model.

\begin{figure*}
	\centering
	\includegraphics[width=2\columnwidth]{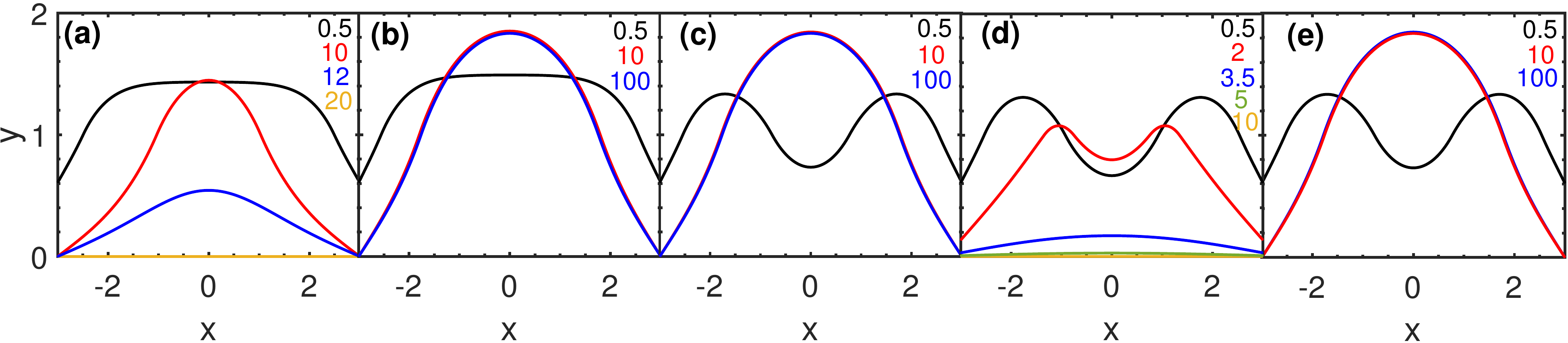}
	\caption{Temperature profiles at different time for different conditions. The increased time is represented by black, red, blue, green, yellow, sequentially. The times (in arbitrary unit) are mentioned by the respective colors. The solution is done with the maximum $t$ of 100. $i_{\rm r}=1.49$. (a) $i=1.45$, below $i_{\rm r}$, stable state is $y=0$ (yellow curve). (b) $i=1.5$, above $i_{\rm r}$, stable NS interface with $y>0$ (blue curve). (c) $i=1.5$ with an initial heat pulse $Q=0.3$ a.u. depicted by the black curve. Stable profile is at $y>0$ (blue curve), and, hence, it remains resistive. (d) With $Q=0.43$, however, stable superconducting state $y=0$ (yellow curve) is back. (e) $Q=0.43$ but the current is increased to $i=1.52$. Stable one is $y>0$, resistive (blue curve).}
	\label{fig:10}
\end{figure*}
\section*{Appendix C: Simulation on the stability of NS interface}
We write the time-dependent equations (\ref{eq:10}) and (\ref{eq:11}) in concise forms as
\begin{align}
	\tau_{\rm th}\frac{\partial y(x,t)}{\partial t}&=\frac{\partial^{2}y ( x,t)}{\partial x^2}-y (x,t)+i^2\Theta[y(x,t)-1]. \label{eq:16}
\end{align}
Here, $x$ is normalized with the thermal healing length $\eta =kd/\alpha$ and $y=\frac{T-T_{\rm b}}{T_{\rm c}-T_{\rm b}}$. We solve the equation for $y(x,t)$ over a length $[-3,3]$ a.u. with increasing time $t$ ($1/\tau_{\rm th}$ a.u.). Using $i_{\rm r}$ expression in the main text and putting $L=6, \eta=1$, the estimated $i_{\rm r}$ comes out to be 1.49. One could see the animated solution of how the profile evolves with time before stabilize. In the following, we show and discuss the snapshot of $y$ at different time $t$.

For $i=1.45<i_{\rm r}$, the system always remains stable at the superconducting state with $T=T_{\rm b}$. As expected, solving the Eq. (\ref{eq:16}) even with an initial condition of $y(x,0)=1$, the stable state is obtained at $y=0$, \mbox{i.e.}, $T=T_{\rm b}$ [see Fig. \ref{fig:10}(a)]. For $i=1.5$ above $i_{\rm r}$, the same initial condition makes a stable temperature profile at $y\neq0$ with an NS interface [see Fig. \ref{fig:10}(b)]; thus, the stable resistive state is formed. However, a different initial condition can lead to the stable superconducting state, verifying the bistable nature above $i_{\rm r}$. To demonstrate the stability of the temperature profile or the NS interface against fluctuations at resistive state, we consider a local heat pulse $Q$ at $x=0$, over a region of width $2w$ and a time of $dt$. Equation (\ref{eq:16}), thus, modifies with an additional term at the left hand side as
\begin{eqnarray}\nonumber
	\tau_{\rm th}\frac{\partial y(x,t)}{\partial t}+\frac{Q}{2wdt}\Theta[w-abs(x)]\Theta[dt-t]=\\
	\frac{\partial^{2}y ( x,t)}{\partial x^2}-y (x,t)+i^2\Theta[y(x,t)-1]. \label{eq:17}
\end{eqnarray}
With $w=0.3$ a.u. and $dt=0.1$ a.u., we solve this equation by varying $Q$ and $i$. The summary of the obtained results are following:\\
1. Figure \ref{fig:10}(c) shows the plots at $i=1.5$. $Q=0.3$ a.u. gives an initial perturbation; however, it is not enough to break the stability.\\
2. Figure \ref{fig:10}(d) shows the plots at $i=1.5$ but with a little higher perturbation $Q=0.43$ a.u. This is just sufficient for the collapse into the superconducting state $y=0$.\\
3. Figure \ref{fig:10}(e) shows the plots with an increased $i=1.52$ for $Q=0.43$ a.u. Again, the stable state is resistive with $y>0$.\\


\begin{thebibliography}{9}
\bibitem{Note} For a short JJ with length $\ell$ much shorter than the superconducting coherence length $\xi$, \mbox{i.e.}, $\ell\ll\xi$, where the supercurrent-phase relation is perfectly sinusoidal, $\beta=3/2$. In the other extreme, \mbox{i.e.}, $\ell\gg\xi$, $\beta=1$. A less $\beta=1.38$ approximately suits the measured $\sigma_{\rm m}$. One could notice that depending on the $\beta$, $\sigma_{\rm t}$ for a particular $I_{\rm c}(T_{\rm b})$ can behave very differently with the temperature.
\end{thebibliography}

\end{document}